\documentclass[runningheads]{llncs}
\usepackage[utf8]{inputenc}
\usepackage[english]{babel}

\usepackage{graphicx} % Required for including pictures
\usepackage{amsfonts}
\usepackage{amsmath}

\usepackage{float} % Allows putting an [H] in \begin{figure} to specify the exact location of the figure
\usepackage{amssymb}
\usepackage{tabularx}
\usepackage{booktabs}
\usepackage[ruled,vlined,linesnumbered,algo2e]{algorithm2e}
	\SetKwInOut{Input}{input}
	\SetKwInOut{Output}{output}

\usepackage{subcaption}

\usepackage{paralist} % inline lists

\ifdefined \VersionLong

\else

\fi
\usepackage{tikz}
\usetikzlibrary{arrows,automata}
\tikzstyle{every node}=[initial text=]
\tikzstyle{location}=[rectangle, rounded corners, minimum size=12pt, draw=black, fill=blue!10, inner sep=2pt]
\tikzstyle{invariant}=[draw=black, dotted, inner sep=1pt] % xshift=1em, 
\usetikzlibrary{positioning}
\tikzset{main node/.style={circle,fill=blue!20,draw,minimum size=0.8cm,inner sep=0pt},
            }

\definecolor{darkblue}{rgb}{0, 0, 0.7}

\usepackage[
		breaklinks  = true,
		colorlinks  = true,
	\ifdefined \VersionWithComments
		pagebackref = true,
	\fi
 		citecolor   = blue!50!blue,
 		linkcolor   = darkblue,
 		urlcolor    = blue!50!green,
	]{hyperref}

\usepackage[capitalise,english,nameinlink]{cleveref} % load after algorithm2e and hyperref
\crefname{line}{\text{line}}{\text{lines}} % to remove the capital

\usepackage{amsthm}

\ifdefined\VersionWithComments
	\usepackage[colorinlistoftodos,textsize=footnotesize]{todonotes}
\else
	\usepackage[disable]{todonotes}
\fi

\ifdefined \VersionWithComments
	\newcommand{\todoinline}[1]{\mbox{}{\color{red}{\textbf{TODO}\ifx#1\\\else:\ \fi #1}}} % here, ``\\'' stands for ``empty''
\else
	\newcommand{\todoinline}[1]{}
\fi

\usepackage{verbatim} % for 'comment'

\usepackage[pagewise,switch]{lineno} % switch, modulo
\definecolor{cof}{RGB}{219,144,71}
\definecolor{pur}{RGB}{186,146,162}
\definecolor{greeo}{RGB}{91,173,69}
\definecolor{greet}{RGB}{52,111,72}
\definecolor{red}{RGB}{210,0,32}

\ifdefined\VersionWithComments
	\usepackage{draftwatermark}
	\SetWatermarkText{draft}
	\SetWatermarkScale{5}
	\SetWatermarkColor[gray]{0.9}
\fi
\ifdefined \VersionWithComments
 	\definecolor{colorok}{RGB}{80,80,150}
\else
	\definecolor{colorok}{RGB}{0,0,0}
\fi

\newcommand{\eg}{\textcolor{colorok}{e.\,g.,}\xspace}
\newcommand{\ie}{\textcolor{colorok}{i.\,e.,}\xspace}

\begin{document}

\title{Constructing invariant tori using guaranteed Euler method
%\todoinline{This is the version with comments. To disable comments, comment out line~3 in the \LaTeX{} source.}
}

%\thanks[footnoteinfo]{Sponsor and financial support acknowledgment
%goes here. Paper titles should be written in uppercase and lowercase
%letters, not all uppercase.}

\author{Jawher Jerray\inst{1} \and 
Laurent Fribourg\inst{2} 
%\and \'Etienne Andr\'e\inst{3}
} 

\institute{Université Sorbonne Paris Nord, LIPN, CNRS, F-93430 Villetaneuse, France \email{jerray@lipn.univ-paris13.fr} \and
Université Paris-Saclay, ENS Paris-Saclay, CNRS, LMF, F-91190 Gif-sur-Yvette, France 
%\and Université de Lorraine, CNRS, Inria, LORIA, F-54000 Nancy, France
}
\maketitle              % typeset the header of the contribution

\begin{abstract}
We show here how, using Euler's integration method and an associated function bounding the error in function of time, one can generate structures closely surrounding the invariant tori of dynamical systems. Such structures are constructed from a finite number of balls of $\mathbb{R}^n$ and encompass the deformations of the tori when small perturbations of the flow of the system occur.

%\keywords{differential equations, \and periodicity \and limit cycle \and stability.}
\end{abstract}

%===============================================================================

\pagestyle{plain}

%\instructions{ADHS 2021: Regular papers: Regular papers can have a length of up to 8 pages at submission. Accepted papers are limited to 6 pages in the conference preprints and on-line proceedings. For papers describing tools, please tick the keyword "Computational tools" upon submission.}

\section{Introduction}
Invariant tori are objects which are omnipresent in physics and intervene in a
 multiplicity of different domains: chemical reactions, population dynamics, electrical circuit theory, electrodynamics, fluid dynamics, $\dots$ 
(see, e.g., \cite{Edoh2000,Rasmussen03}).
These tori are (positively) {\em invariant} in the sense that all the orbits 
lying on their surface  at $t=t_0$ remain on them at all 
subsequent time $t\geq t_0$.

The topology of tori conveys important 
information. In order to understand it, one 
introduces a ``continuation'' parameter (say $\mu$) in the equations of the dynamical system,
a simple basic case corresponding to $\mu=0$. One then progressively
make $\mu$ vary, and observe the change of topology of the torus.
Roughly speaking, a torus appears when, in a Poincar\'e section, 
a stable fixed-point $M$ becomes unstable while
an invariant closed curve (``circle'')  $L$ appears around~$M$.
In the full space, $M$
corresponds to a {\em repulsive circle} $C$ of the system, and~$L$
to an attractive {\em  invariant torus} ${\cal T}$.
Further variations of $\mu$ lead to the deformation of ${\cal T}$
until a ``torus bifurcation'' occurs.
When $\mu$ is still modified, the solutions of the system
become ``aperiodic'' and a phenomenon of {\em chaos} appears.

There are basically three  kinds of methods of numerical analysis that exploit
this mechanism of parameter continuation:  
partial differential equation  \cite{Dieci-Bader94,Trummer2000}, graph 
transform \cite{Broer02,Reichelt2000,Schilder2005}
and orthogonality methods \cite{Edoh2000,Rasmussen-Dieci08}. 
Their respective advantages and disadvantages are analyzed,
basically from a computational efficiency point of view,
in, e.g., \cite{Edoh2000,Rasmussen03}.
From a formal point of view, 
all the methods are incomplete because they focus on a {\em discretization} of 
the continuous dynamical system, but do not take the associated errors in consideration, or, at best, evaluate them modulo unknown constants
(see, e.g.,~\cite{Garay2001,Hairer99,Reichelt2000}). 

On the other hand, Capinski and co-authors recently developed
a {\em guaranteed} computer assisted 
method of proof for attractive invariant tori (see \cite{Capinski2020}). 
%Their analysis is formulated in terms of geometric hypotheses checked with {\em rigorous} computational techniques. 
They obtain an {\em outer approximation} of
the torus via covering by polygons. 
Their implementation is based on the validated integrators developed 
by Wilczak and Zgliczynski \cite{Wilczak2007}.
We follow a similar approach, but rely here
on Euler's integration method 
associated to an  error function~$\delta(t)$
that bounds, at time $t$,
the distance between the numerical and the 
exact solutions (see \cite{SNR17}).  
We are thus able to generate a {\em finite}  number of
$n$-dimensional balls of radius $\delta(t)$ for bounded values of $t$, 
which encompass the torus. 
The set of balls is itself invariant and continues to contain the 
torus when the latter deforms under small variations of~$\mu$. 
This approach extends our previous work~\cite{ADHS21}, which 
was limited to the determination of invariant circles. 
%to higher dimensional invariant tori.
%\footnote{We know that, in the case of a two-dimensional torus bounded by two closed trajectories (``circles'') \cite{Denjoy1932}, the trajectories typically spiral away from one (the repulsive one) and spirals towards the other (the attractive one) in the opposite direction. Such phenomena also often occur in higher dimensions.}

%%%%%%%%%%%%%%%%%%%%

%\paragraph{Plan of the paper}

\section{Preliminaries}\label{sec:method}
\subsection{Euler's method and error bounds}\label{ss:Euler}
%, and 
%we suppose that the control law ${\bf u}(\cdot)$ is a {\em piecewise-constant} function, which takes its
%values on a {\em finite} set~$U$, called ``set of modes''.
%Given $u\in U$, l
Let us consider the differential system: 
%$$\frac{dy}{dt}=\sigma{\cal L}_h y+\sigma\varphi_h(t,u)+f(t,y)$$
%by:
$$\dot{x}(t)=f(x(t)),$$
with states $x(t)\in\mathbb{R}^n$.
%
%$f_u(y(t))$ stands for $f({\bf u}(t),y(t))$ with ${\bf u}(t)=u$ for $t\in[0,\tau]$, and $y(r)\in\mathbb{R}^n$ denotes the state of the system at time $t$.
%
%{\cal L}_hy(t,x)+\varphi_h(t,u)+f_h(t,y).$$
%For $t\in [0,\tau]$ and $u\in U$,
We will  use $x(t;x_0)$ (or sometimes just $x(t)$) to denote  the exact continuous solution %~$x$ %:[0,\tau]\times \Omega_h\rightarrow [0,1]^n$ 
of the system at time~$t$,
for a given initial condition~$x_0$.
We use $\tilde{x}(t;y_0)$
(or just $\tilde{x}(t)$) to denote Euler's approximate value 
of $x(t;y_0)$
(defined by $\tilde{x}(t;y_0)=y_0+tf(y_0)$  for $t\in [0,\tau]$, where $\tau$ is the integration time-step).

We suppose that we know a bounded region
${\cal S}\subset \mathbb{R}^n$ containing
the solutions of the system for a set of initial conditions $B_0$
and  a certain amount of time.
%
%\subsection{Error bounds}\label{ss:error}
%We first consider the ODE:
%$\frac{dy}{dt}=f(y,{\bf 0})$, and 
We now give an upper bound to the error between
the exact solution of the ODE and
its Euler approximation on ${\cal S}$ (see~\cite{SNR17}).

%%%\mbox{ }

\begin{definition}\label{def:delta}
Let $\varepsilon$ be a given positive constant. Let us define, for $t\in [0,\tau]$,
$\delta_{\varepsilon}(t)$ as follows:\\
%\begin{itemize}
%\item  
$\mbox{if } \lambda <0:$
$$\delta_{\varepsilon}(t)=\left(\varepsilon^2 e^{\lambda t}+
 \frac{C^2}{\lambda^2}\left(t^2+\frac{2 t}{\lambda}+\frac{2}{\lambda^2}\left(1- e^{\lambda t} \right)\right)\right)^{\frac{1}{2}}$$
%
%\item 
$\mbox{if } \lambda = 0:$
$$\delta_{\varepsilon}(t)= \left( \varepsilon^2 e^{t} + C^2 (- t^2 - 2t + 2 (e^t - 1)) \right)^{\frac{1}{2}}$$
%\item if $\lambda^a = 0:$
%$$\delta_t= \frac{C t^2}{2} + \delta$$
%
$\mbox{if } \lambda > 0:$
%\begin{dmath*}
$$\delta_{\varepsilon}(t)=\left(\varepsilon^2 e^{3\lambda t}+ 
\frac{C^2}{3\lambda^2}\left(-t^2-\frac{2t}{3\lambda}+\frac{2}{9\lambda^2}
\left(e^{3\lambda t}-1\right)\right)\right)^{\frac{1}{2}}$$
%\end{dmath*}
%
%\end{itemize}
where $C$ and $\lambda$ are real constants specific to function $f$,
defined as follows:
$$C=\sup_{y\in {\cal S}} L\|f(y)\|,$$
where $L$ denotes the Lipschitz constant for $f$, and
$\lambda$ is the ``one-sided Lipschitz constant'' (or ``logarithmic Lipschitz constant'' \cite{Aminzare-Sontag}) associated to $f$, \ie{} the 
minimal constant such that, for all $y_1,y_2\in {\cal S}$:
$$\langle f(y_1)-f(y_2), y_1-y_2\rangle \leq \lambda\|y_1-y_2\|^2,\ \ \ \ \ (H0)$$
where $\langle\cdot,\cdot\rangle$ denotes the scalar product of two vectors
of ${\cal S}$ and $\|\cdot\|$ the Euclidean norm.
\end{definition}

%%%\mbox{}

The constant $\lambda$ can be computed using a nonlinear optimization solver (\eg{} CPLEX~\cite{cplex2009v12}) or using the Jacobian matrix of $f$ (see, \eg{}~\cite{Aminzare-Sontag}).

\subsection{Systems with bounded uncertainty}

Let us now show how the method extends to 
systems with ``disturbance'' or ``bounded uncertainty''.
%\section{Robustness against bounded perturbations}\label{appendixun}
A differential system $\Sigma_{{\cal W}}$ with bounded uncertainty is of the form 
$$\dot{x}(t)=f(x(t),w(t)),$$
with $t\in\mathbb{R}^n_{\geq 0}$,
states $x(t)\in\mathbb{R}^n$, and uncertainty $w(t)\in{\cal W}\subset \mathbb{R}^n$ (${\cal W}$ is compact, \ie{} closed and bounded).
We assume that any possible disturbance trajectory is bounded at any
point in time in the compact set $W$. We denote this by
$w\in{\cal W}$, which is a shorthand for
$w(t)\in {\cal W},\forall t\geq 0$.
See~\cite{SchurmannA17,SchurmannA17b} for details. 
%
%\footnote{We denote this by $w(\cdot)\in{\cal W}$, which is a shorthand for $w(t)\in {\cal W}, \forall t\in [0,\tau]$.}. 
%The same shorthand is also used for state and input trajectories.} 
%\footnote{LF???:Without loss of understanding, we still use $x(t;y_0)$, or just $x(t)$, to denote the solution of $\frac{dx(t)}{dt}=f(x(t),w(t))$ with initial condition $y_0$. We also use $\tilde{x}(t; x_0)$, or just $\tilde{x}(t)$, to denote the approximate Euler solution  of the system $\frac{dx(t)}{dt}=f(x(t),0)$ {\em without uncertainty} (\ie{} when ${\cal W}=0$), with initial condition $x_0$.}
%
%The solution of an undisturbed system  (\ie{} with ${\cal W}=0$) is denoted by $\phi_{u}(t;y_0,0)$. 
%
%We now suppose that ${\cal S}$ is Euler-invariant for the
%system with {\em null perturbation}, \ie{} for all $y\in {\cal S}$,
%there exists $u$ such that $\tilde{Y}_{y,{\bf 0}}(\tau)\in {\cal S}$.
%\footnote{not restrictive??? compare to Assumption 5.1}
%and $\phi_\pi(\tau;y_0,0)\in {\cal S}$.
%
We now suppose (see~\cite{AdrienRP17}) that
there exist constants $\lambda\in\mathbb{R}$ 
and $\gamma\in\mathbb{R}_{\geq 0}$ such that,
for all $y_1,y_2\in {\cal S}$ and $w_1,w_2\in {\cal W}$:\\

$\langle f(y_1,w_1)-f(y_2,w_2), y_1-y_2\rangle$

$\leq 
\lambda\|y_1-y_2\|^2 + \gamma \|y_1-y_2\|\|w_1-w_2\|\ \ \ \ \ (H1).$\\
%where $\langle\cdot,\cdot\rangle$ denotes the scalar product of two vectors of ${\cal S}$.
\\
This formula can be seen as a generalization of (H0) (see \cref{ss:Euler}). 
Recall that $\lambda$ has to be computed in the absence of uncertainty
(${\cal W}=0$). The additional constant $\gamma$ is used for taking into account
the uncertainty $w$.
Given~$\lambda$, the constant $\gamma$ can be computed itself
using a nonlinear optimization solver (\eg{} CPLEX~\cite{cplex2009v12}).
%\footnote{LF: Note that the notion of contraction (often used in the literature~\cite{ManchesterCDC13,Aminzare-Sontag}) corresponds to the case where $\lambda<0$???}
%, but we do not need this assumption here ($\lambda$ can be positive, at least locally). 
%
We now give a property
%a version of \cref{prop:basic} with bounded uncertainty $w(\cdot)\in{\cal W}$, 
originally proved in~\cite{AdrienRP17}.

%%%\mbox{ }

\begin{proposition}\label{prop:1bis}\cite{AdrienRP17}
Consider
a system $\Sigma_{{\cal W}}$ with bounded uncertainty
of the form $\dot{x}(t)=f(x(t),w(t))$
satisfying (H1).
%and, for all $u\in U$:
%
%$e>gw/2  \wedge  e  < sqrt(2) C/l^2  \wedge e < 2/5^{1/4}  sqrt(C)/l^2)
%\wedge t*< sqrt(3)   e |l| / C$, 
%
%where $t^*$ is the positive real
%root of equation
%$F(t)= Pt^2 + Qt +R = 0$
%		with $ P = –(d|l|^3/6), Q = (a+dl^2/2), R = (b-d|l|)$.
%
Consider a point~$x_0\in {\cal S}$ and a point $y_0\in B(x_0,\varepsilon)$\footnote{As usual, $B(x_0,\varepsilon)$ denotes the ball of center $x_0$ and radius $\varepsilon$  defined by $B(x_0,\varepsilon) :=\{x'\in {\cal S}\ |\ \|x_0-x'\|\leq \varepsilon\}$.}.
Let $x(t;y_0)$ be the exact solution of $\Sigma_{{\cal W}}$
with bounded uncertainty ${\cal W}$
and initial condition $y_0$,
and $\tilde{x}(t;x_0)$ the Euler approximate solution of the system
$\Sigma_0: \dot{x}(t)=f(x(t),0)$ {\em without uncertainty}
(${\cal W}=0$) with initial condition $x_0$.
%of $\varepsilon$-representative $z_0\in {\cal X}$.
% Let us denote by $\lambda_{u}$ the greatest eigenvalue 
% of $\frac{A_{u} + A_{u}^\top}{2}$ for all $u \in U_1$.
% Suppose sub-system 1 verifies $$f_{\sigma_1} (x_1,x_2) = A x_1 + B x_2 + u_{\sigma_1}$$
We have,
for all $w \in {\cal W}$ and $t\in[0,\tau]$: %and any $\sigma_2 \in U_2$:
%
%$$\phi_j(t;y_0)\in {\cal B}(\tilde{y}_0-t f_j(\tilde{y}_0), \gamma)$$
$$\|x(t;y_0)-\tilde{x}(t;x_0)\|\leq \delta_{\varepsilon,{\cal W}}(t)$$
with %$y_0 = \begin{pmatrix}y_0 \\y_2_0\end{pmatrix}$ and
\begin{itemize}
\item if $\lambda <0$,
\begin{multline}
 \delta_{\varepsilon,{\cal W}}(t) = 
%  \frac{1}{\lambda_{u}^{3/2}} 
 \left( \frac{C^2}{-\lambda^4} \left( - \lambda^2 t^2 - 2 \lambda t + 2 e^{\lambda t} - 2 \right) \right.   \\
 + \left. \frac{1}{\lambda^2} \left( \frac{C \gamma |{\cal W}|}{-\lambda} \left( - \lambda t + e^{\lambda t} -1 \right) \right. \right.  \\ + \left. \left. \lambda \left( \frac{\gamma^2 (|{\cal W} |/2)^2}{-\lambda} (e^{\lambda t } - 1) + \lambda \varepsilon^2 e^{\lambda t}  \right) \right)  \right)^{1/2}
\end{multline}
where $|{\cal W}|$ denotes the maximum distance between two elements of
${\cal W}$.
%\begin{itemize}
% \item if $\lambda^1_{j_1} <0$,
%\begin{multline}
% \delta_{j_1} (t) = 
%  \frac{1}{\lambda^1_{j_1}^{3/2}} 
% \left( \frac{(C_{j_1}^1)^2}{-(\lambda^1_{j_1})^4} \left( - (\lambda^1_{j_1})^2 t^2 - 2 \lambda^1_{j_1} t + 2 e^{\lambda^1_{j_1} t} - 2 \right) \right.   \\
% + \left. \frac{1}{(\lambda^1_{j_1})^2} \left( \frac{C_{j_1}^1 \gamma^1_{j_1} |S_2|}{-\lambda^1_{j_1}} \left( - \lambda^1_{j_1} t + e^{\lambda^1_{j_1} t} -1 \right) \right. \right.  \\ + \left. \left. \lambda^1_{j_1} \left( \frac{(\gamma^1_{j_1} )^2 (|S_2 |/2)^2}{-\lambda^1_{j_1}} ( e^{\lambda^1_{j_1} t } - 1) + \lambda^1_{j_1} \delta^2 e^{\lambda^1_{j_1} t}  \right) \right)  \right)^{1/2}
%\end{multline}
\item if $\lambda >0$,
\begin{multline}
 \delta_{\varepsilon,{\cal W}}(t) = \frac{1}{(3\lambda)^{3/2}} \left( \frac{C^2}{\lambda} \left( - 9\lambda^2 t^2 - 6\lambda t + 2 e^{3\lambda t} - 2 \right) \right.   \\
 + \left. 3\lambda \left( \frac{C  \gamma |{\cal W}|}{\lambda} \left( - 3\lambda t + e^{3\lambda t} -1 \right) \right. \right.  \\
 + \left. \left. 3\lambda \left( \frac{\gamma^2 (|{\cal W} |/2)^2}{\lambda} ( e^{3\lambda t } - 1) + 3\lambda \varepsilon^2 e^{3\lambda t}  \right) \right)  \right)^{1/2}
\end{multline}
\item if $\lambda = 0$, 
\begin{multline}
 \delta_{\varepsilon,{\cal W}}(t)= 
%  \frac{1}{\lambda^{3/2}} 
 \left( {C^2} \left( -  t^2 - 2  t + 2 e^{ t} - 2 \right) \right.   \\
 + \left.  \left( {C \gamma |{\cal W}|} \left( -  t + e^{ t} -1 \right) \right. \right.  \\ + \left. \left.  \left({\gamma^2 (|{\cal W} |/2)^2} ( e^{ t } - 1) +  \varepsilon^2 e^{ t}  \right) \right)  \right)^{1/2}
\end{multline}
\end{itemize}
\end{proposition}
%

%%%\mbox{ }

We will sometimes write $\delta_{{\cal W}}(t)$ instead of
$\delta_{\varepsilon,{\cal W}}(t)$.
%Let $\delta_{{\cal W}}(t)\equiv  \delta_{\varepsilon,{\cal W}}(t)$
%Let $B(t)\equiv B(\tilde{Y}_{x_0,{\bf 0}}(t),\delta_{\varepsilon,{\cal W}}(t))$.
%with $\tilde{y}_{{\cal W}}(0)=z_0$.
%\cref{prop:1bis} expresses that, for $t\in[0,\tau]$, the ``tube'' 
%$\bigcup_{t\geq 0}B(t)$ contains all the 
%solutions $Y_{y_0,{\cal W}}(t)$ with $\|y_0-x_0\|\leq \varepsilon$,
%and is therefore {\em robustely (positive) invariant}.

%%%\mbox{ }

Actually, we will not compute $\lambda$ (resp. $\gamma$) {\em globally} for ${\cal S}$, but will
decompose ${\cal S}$ into a set of subregions~$\{{\cal S}_i\}_{i=1,\dots,k}$
with $k\tau=T$, where ${\cal S}_i$ 
is an appropriate subregion of ${\cal S}$ enclosing
the states of the system state during the interval
of time $[(i-1)\tau, i\tau]$. Instead of a global upperbound of $\lambda$ satisfying $(H1)$ on ${\cal S}$, we will compute a {\em local} upperbound
$\lambda_i$  (resp. $\gamma_i$) of $\lambda$ 
(resp. $\gamma$) on  each subregion~${\cal S}_i$ ($1\leq i\leq k$).
\cref{prop:1bis} extends naturally in this context.
%This will allow us to deal with systems that are ``globally contractive'' but possibly ``locally expansive'' (when $\lambda_i>0$ for some  $1\leq i\leq k$). 
%The property of convergence of two trajectories towards each other is preserved.
%Instead of computing them globally
%for ${\cal S}$, it is advantageous to compute $\lambda$ and $\gamma$ {\em locally} depending on the subregion of ${\cal S}$ occupied by the system state during a considered interval of time.
%\section{Construction of Invariant Structures around the Tori}
\section{Constructing Invariant Stuctures Around Tori}
%\section{Invariant  lassos}
%\subsection{Determination of invariant tubes in form of lasso}
Consider a differential system $\Sigma_{{\cal W}}: \dot{x}=f(x,w)$
with $w\in{\cal W}$, an initial point $x_0\in\mathbb{R}^n$, a real $\varepsilon>0$ and a ball $B_0=B(x_0,\varepsilon)$. Let
${\cal B}_{\varepsilon,{\cal W}}(t)$ denote $B(\tilde{x}(t), \delta_{\varepsilon,{\cal W}}(t))$
where $\tilde{x}(t)$ is the Euler approximate solution of the system
without uncertainty and initial condition $x_0$\footnote{Note that ${\cal B}_{\varepsilon,{\cal W}}(0)=B_0$ because
$\tilde{x}(0)=x_0$ and $\delta_{\varepsilon,{\cal W}}(0)=\varepsilon$.}.
It follows from \cref{prop:1bis} that $\bigcup_{t\geq 0} {\cal B}_{\varepsilon,{\cal W}}(t)$ 
is an invariant set containing $B_0$.
We can make a stroboscopic map of this invariant.
by considering periodically the set ${\cal B}_{\varepsilon,{\cal W}}(t)$ at the moments
$t=0,T,2T$, etc., with $T=k\tau$ for some $k$ 
($\tau$ is the time-step used in Euler's method). The value of $T$ is an estimate of the exact period $T^*$ of the system.

If moreover, we can find an integer $i\geq 0$ such that
${\cal B}_{\varepsilon,{\cal W}}((i+1)T)\subseteq {\cal B}_{\varepsilon,{\cal W}}(iT)$, then
%la suite $\{B(jT)\}_{j=i,i+1,\dots}$ est decroissante pour la relation d'ordre 
%d'inclusion des ensembles, et 
we have
${\cal B}_{\varepsilon,{\cal W}}(iT)=\bigcup_{j=i,i+1,\dots}{\cal B}_{\varepsilon,{\cal W}}(jT)$
%$\bigcup_{t\geq 0}B(t)=\bigcup_{t\in[0,(i+1)T]}B(t)$ et
and $\bigcup_{t\in[0,(i+1)T]}{\cal B}_{\varepsilon,{\cal W}}(t)=\bigcup_{t\geq 0}{\cal B}_{\varepsilon,{\cal W}}(t)$.
The set 
$\bigcup_{t\in[0,(i+1)T]}{\cal B}_{\varepsilon,{\cal W}}(t)$ %(abbreviated to ${\cal B}_{\varepsilon,{\cal W}}([0,(i+1)T])$)
is thus a {\em bounded invariant} 
which contains all the solutions $x(t)$
starting at $B_0$, for $t\in[0,\infty)$.
%In the phase space, this bounded invariant has a ``torus'' shape.
We have:

%%%\mbox{ }

\begin{proposition}\label{prop:inclusionbis}\cite{ADHS21}
Consider a system $\Sigma_{{\cal W}}: \dot{x}=f(x,w)$ with uncertainty $w\in{\cal W}$ 
satisfying $(H1)$, and
a set of initial conditions $B_0\equiv B(x_0,\varepsilon)$. 
Suppose that there exist $T>0$ (with $T=k\tau$ for some $k\in\mathbb{N}$) 
and $i\in\mathbb{N}$ such that

(*):\ \ \ \  ${\cal B}_{\varepsilon,{\cal W}}((i+1)T) \subseteq {\cal B}_{\varepsilon,{\cal W}}(iT)$. \\
Then we have:
\begin{enumerate}
\item $\bigcup_{t\in[0,(i+1)T]} {\cal B}_{\varepsilon,{\cal W}}(t)$ is a
compact (i.e., bounded and closed) invariant set containing, for $t\in[0,\infty)$, all the solutions $x(t)$ of $\Sigma_{{\cal W}}$  with initial condition in~$B_0$.
\item  The subset $\bigcup_{t\in[iT,(i+1)T]}{\cal B}_{\epsilon,{\cal W}}(t)$ contains an attractive circle (or ``stable limit cycle'')
of the system $\Sigma_0$ without uncertainty ($w=0$).
\end{enumerate}
\end{proposition}
%\begin{proof}
%(Sketch)
%\end{proof}

%%%\mbox{ }

%\begin{remark}
\cref{prop:inclusionbis} states
that the invariant set $\bigcup_{t\in[0,(i+1)T]}{\cal B}_{\varepsilon,{\cal W}}(t)$ 
is an $n$-dimensional tube having the form of a ``lasso''composed of 
a linear part $\bigcup_{t\in[0,iT]} {\cal B}_{\varepsilon,{\cal W}}(t)$
connected to a looping part $\bigcup_{t\in[iT,(i+1)T]} {\cal B}_{\varepsilon,{\cal W}}(t)$. Besides,
the looping part encloses a  1-dimensional attractive circle.
Since ${\cal B}_{\varepsilon,{\cal W}}(t)$ is a ball of $\mathbb{R}^n$
(of radius $\delta_{\varepsilon,{\cal W}}(t)$),
a lasso is constructed from a {\em finite} number 
({\em viz.}, $(i+1)\times k$) of balls.

Given $\varepsilon, {\cal W}, \tau$, $T=k\tau$,
the lasso: $\bigcup_{t\in[0,(i+1)T]}{\cal B}_{\varepsilon,{\cal W}}(t)$ is uniquely determined by the center $x_0\in\mathbb{R}^n$ of the initial ball $B_0=B(x_0,\varepsilon)$
and by $i_0$, the integer such that (*) holds.
%${\cal B}_{\varepsilon,{\cal W}}((i_0+1)T)\subseteq {\cal B}_{\varepsilon,{\cal W}}(i_0T)$ holds.
We call $x_0$ the {\em source point} of the lasso, and $B_0=B(x_0,\varepsilon)$ the source ball.
We will denote such a lasso by ${\cal L}(x_0,i_0)$ or more simply
by ${\cal L}(x_0)$, where $i_0$ is left implicit.
%\end{remark}
%\begin{remark}
Note that the invariance property of a lasso ${\cal L}(x_0)$
%$\bigcup_{t\in[0,(i+1)T]}{\cal B}_{\varepsilon,{\cal W}}(t)$ 
is {\em robust}: the invariance persists even in presence of a bounded perturbation $w\in{\cal W}$ of the dynamical system.
%\end{remark}

The implementation of the construction of lassos has been done in Python
% 	by Jawher Jerray.
and corresponds to a program of
around 500 lines.
The source code is available at
	\href{https://lipn.univ-paris13.fr/~jerray/orbitador/}{\nolinkurl{lipn.univ-paris13.fr/~jerray/orbitador/}}.
In the experiments below, the program runs on a 2.80 GHz Intel Core i7-4810MQ CPU with 8\,GiB of memory.
Given $x_0$, one searches for values of $\tau,\varepsilon,{\cal W},T$ at hand
(by trial and error) so that inclusion
(*) can be successfully verified by the program.
%in order to make the program successfully verify ${\cal B}_{\varepsilon,{\cal W}}((i+1)T)\subset  {\cal B}_{\varepsilon,{\cal W}}(iT)$ for some $i\in\mathbb{N}$. 

\begin{example}\label{ex:vdp0}
Consider the forced Van der Pol (VdP) system $\Sigma_{{\cal W}}$ with
initial condition in $B_0=B(x_0,\varepsilon)$
for some $x_0\in\mathbb{R}^3$ and $\varepsilon>0$ (adapted from~\cite{Rasmussen03}).
%???\footnote{Note that a sign error in the differential system has been corrected here???};cf. \cite{Reichelt2000}???):

$\dot{x_1}=\frac{x_1(\sqrt{x_1^2+x_2^2}-3)}{\sqrt{x_1^2+x_2^2}}(\mu -(\sqrt{x_1^2+x_2^2}-3)^2-x_3^2)
-\frac{x_2^2+x_1x_3}{\sqrt{x_1^2+x_2^2}})+w$

$\dot{x_2}=\frac{x_2(\sqrt{x_1^2+x_2^2}-3)}{\sqrt{x_1^2+x_2^2}}(\mu -(\sqrt{x_1^2+x_2^2}-3)^2-x_3^2)
+\frac{x_1x_2-x_2x_3}{\sqrt{x_1^2+x_2^2}})+w$

$\dot{x_3} =(\sqrt{x_1^2+x_2^2}-3)+\mu x_3 - x_3((\sqrt{x_1^2+x_2^2}-3)^2+x_3^2)+w$\\
with a parameter $\mu$ that controls the periodic forcing term
and a bounded perturbation $w\in{\cal W}$.
Here $\mu=1$ and ${\cal W}=[-0.001,0.001]$.
%\footnote{Without forcing ($\mu=0$) we have an oscillator with an attractive period orbit.}.
%We have $\mu=\mu_0+w$\footnote{LF: Est-ce bien sur $\mu$ que $w$ intervient???}\jj{Dans cette exemple, $w$ est  une perturbation additive qui s'applique sur les états (càd: $\dot{x_1}+w$, $\dot{x_2}+w$ ...)} with $\mu_0=1$ and 
%
%
Let the time-step be equal to $\tau = 10^{-3}$ and the radius of the initial
ball around the source points be $\varepsilon=0.05$.
Let $T=6.283$ be used as an approximation of the exact period $T^*=2\pi$ of the system.
Let $X(t) := (x_1(t), x_2(t), x_3(t))$ with source point
$X(0) := (4, -10^{-3}, -4.8985872 \cdot 10^{-16})$ and $\delta_{{\cal W}}(0)= \varepsilon=0.05$. We have:

$X(T)= (3.96480714, -5.31384851 \cdot 10^{-1}, -1.78122434 \cdot 10^{-4}), \delta_{{\cal W}}(T)= 0.009369013554590614$

$X(2T)= (-3.99399126, -2.23716163 \cdot 10^{-1}, -1.12718456 \cdot 10^{-3}), \delta_{{\cal W}}(2T)= 0.013528832294010595$

$X(3T)= (-4.00024909, -4.16869048 \cdot 10^{-4}, -1.31261670 \cdot 10^{-3}), \delta_{{\cal W}}(3T)= 0.008339289838071407$

$X(4T)= (-4.00024885, -7.76179315 \cdot 10^{-7}, -1.49692259 \cdot 10^{-3}), \delta_{{\cal W}}(4T)= 0.008181686420182348$

$X(5T)= (-4.00024856, -1.44518842 \cdot 10^{-9}, -1.68122843 \cdot 10^{-3}), \delta_{{\cal W}}(5T)= 0.008088285030977036$

$X(6T)= (-4.00024823, -2.69083383 \cdot 10^{-12}, -1.86553421 \cdot 10^{-3}), \delta_{{\cal W}}(6T)= 0.008001319005309636$

$X(7T)= (-4.00024787, -5.01013330 \cdot 10^{-15}, -2.04983993 \cdot 10^{-3}), \delta_{{\cal W}}(7T)= 0.008408806943539475$

$X(8T)= (-4.00024747, -9.32849705 \cdot 10^{-18}, -2.23414558 \cdot 10^{-3}), \delta_{{\cal W}}(8T)= 0.007976450475139826$.

%$X(9T)= (-4.00024705, -1.73689703 \cdot 10^{-20}, -2.41845115 \cdot 10^{-3}), \delta_{{\cal W}}(9T)= 0.008430180016494303$

%$X(10T)= (-4.00024658, -3.23397355 \cdot 10^{-23}, -2.60275665 \cdot 10^{-3}), \delta_{{\cal W}}(10T)= 0.007978254320651546$.

We have: ${\cal B}_{{\cal W}}(8T)\subset {\cal B}_{{\cal W}}(7T)$,
i.e.:  ${\cal B}_{{\cal W}}((i_0+1)T)\subset {\cal B}_{{\cal W}}(i_0T)$ 
for $i_0=7$.%\footnote{NB: We have also $B(10T)\subset B(9T)$???}.
 The computation takes 1038 seconds of CPU time.
%La somme de lambda pendant une période est égale à -6289.0986<0. 
See \cref{fig:lasso}. 
\begin{figure}[h!]
\centering
\includegraphics[scale=0.4]{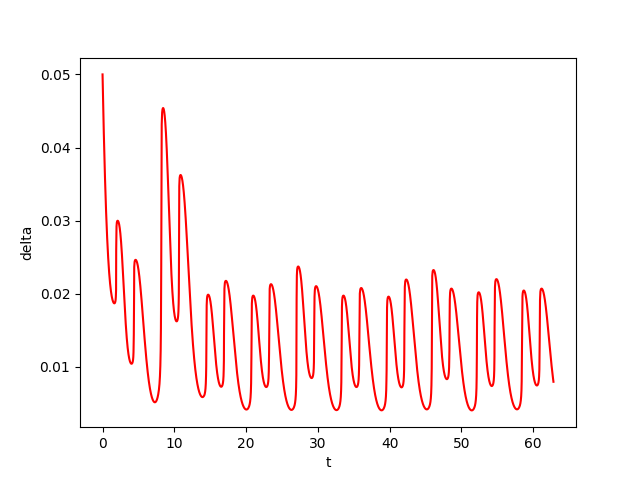}
%two-periodic-orbits-system-delta-dt=1-on-1000,0-tf=62,83-w=0,001-epsilon=0,05-7.png
%\includegraphics[scale=0.4]{figures/vanderpol-single-orbit-u1-u2-dt=1-on-1000,0-tf=33,73-w=1-epsilon=0,5_4-3.png}
\includegraphics[scale=0.4]{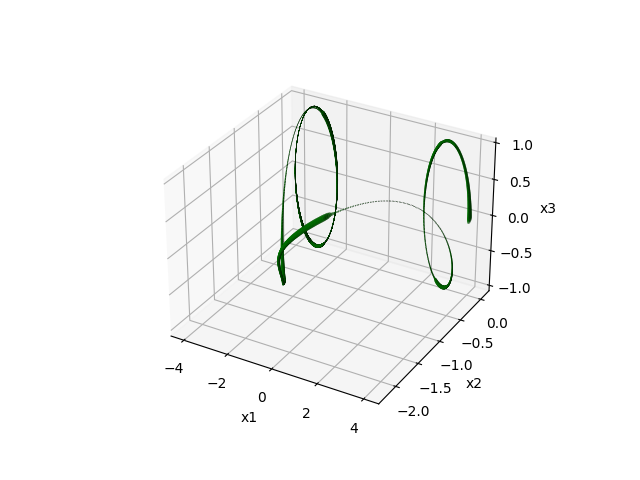}
\caption{%An invariant lasso of the forced VdP system.
{\em Forced VdP}. Top: the function $\delta_{{\cal W}}(t)$ giving the evolution of the radius of a
lasso ball.
Bottom: the corresponding invariant lasso.}
\label{fig:lasso}
\end{figure}
An analogous computation of lassos for 3 other source points takes 4052 seconds.
The 4 lassos are depicted together on \cref{fig:lasso4}.

\begin{figure}[h!]
\centering
\includegraphics[scale=0.4]{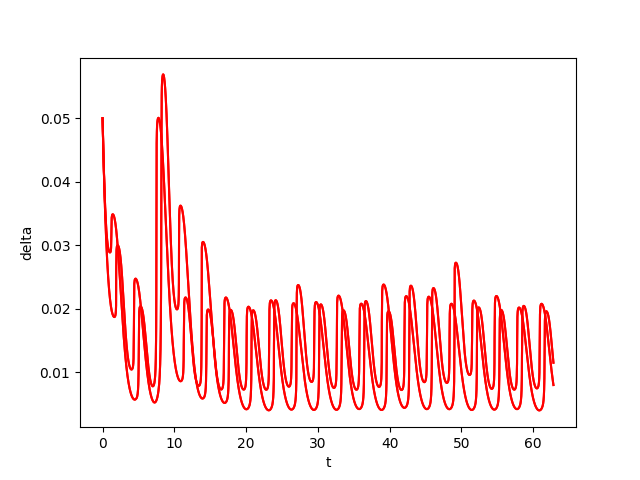}
\includegraphics[scale=0.4]{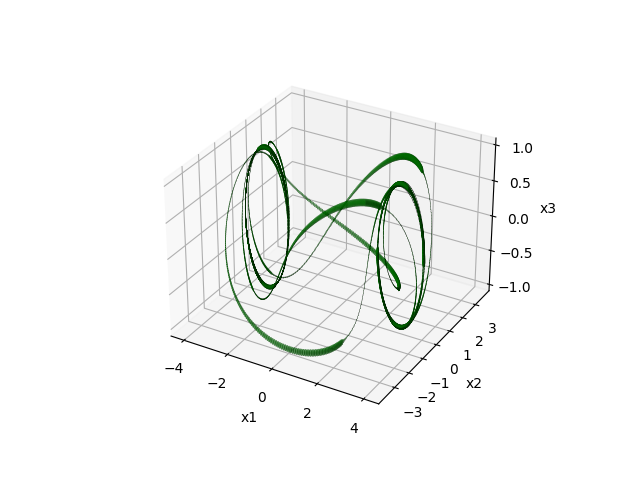}
\caption{{\em Forced VdP}. Top: the function $\delta_{{\cal W}}(t)$ giving the evolution of the radius of 4 lasso balls. Bottom: the 4 corresponding
invariant lassos.}
\label{fig:lasso4}
\end{figure}
\end{example}
%\section{Constructing Invariant Stuctures Around Tori}
%By generating lassos corresponding to different source points, we form a juxtaposition of a (finite) number of balls of $\mathbb{R}^n$ that will progressively encompass the sought invariant torus. 

Given a closed orbit (``circle'') $C$, and a union ${\cal R}$ of
balls of radius $\varepsilon>0$, we say that  {\em ${\cal R}$ isolates $C$}  if there exists
$\alpha>0$ such that:

(**)\ \ \ Any continuous curve containing a point of $C$ and a point located at distance 
$\alpha$ from $C$, also contains a point of ${\cal R}$.\\ 
We say that {\em ${\cal R}$ is at distance $\alpha_0>0$ of $C$}, where $\alpha_0$ is the greatest
$\alpha$ satisfying property~(**).

Let ${\cal T}$ be a torus of repulsive circle $C$, and
${\cal M}$ a set of lassos. We say that ${\cal M}$ {\em covers  
${\cal T}$ (besides the $\alpha_0$-neighborhood of
$C$}), if all orbit ${\cal O}$ on ${\cal T}$ starting
at a distance greater than $\alpha_0$ from $C$ %\footnote{i.e., the distance of all point of ${\cal O}$ to $C_1$ is at greater than $\alpha_0$.},
is contained in a lasso of ${\cal M}$. We have:

\begin{theorem}\label{th:main}
Let ${\cal T}$ be a torus of repulsive circle $C$, and ${\cal R}$ a union of balls of $\mathbb{R}^n$
isolating $C$ at distance $\alpha_0$.
The set of lassos ${\cal M}$ having the balls of ${\cal R}$ as source balls, covers  
${\cal T}$ (besides the $\alpha_0$-neighborhood of $C$).
Furthermore, ${\cal M}$ continues to cover 
${\cal T}$ for 
a bounded perturbation $w\in{\cal W}$ of the dynamical system. 
%of the continuation parameter $\mu_0$ occurs (i.e., $\mu=\mu_0+w$ with $w\in{\cal W}$).

\end{theorem}

The proof is based on the fact that,
by \cref{prop:inclusionbis}, each lasso of ${\cal M}$ connects 
its source ball to an attractive circle. (The full proof will be given in
the long version of this paper.)
Note that the application of \cref{th:main} requires the prior estimate
of the location of the torus repulsive circle $C$. Actually, as seen in the forthcoming examples, taking a subset ${\cal R}'$ of ${\cal R}$ as 
source balls,
%in the vicinity of $C_1$ 
even if ${\cal R}'$ does not isolate $C$ ``completely'', suffices to provide
useful information on~${\cal T}$.

\begin{example}\label{ex:forced_vdp}
For the system of \cref{ex:vdp0}, we generate
100 lassos which
(partially) cover the
invariant torus of the system, as depicted on \cref{fig:lasso50}.
The choice of the 100 source points is as follows. One knows (see \cite{Rasmussen03})
that the system has, in the $x_2$-$x_3$ plane, a {\em repulsive} invariant circle $C$ of centre $(3,0,0)$ and radius~$1$. We thus take 100 source points distributed in the vicinity of the circumference of~$C$.%\footnote{As mentioned above, the prior estimate of the position of the repulsive invariant circle is  useful to find a suitable distribution of source points that entails the covering of the torus with a minimum number of lassos.} 
 The same values of $\varepsilon,\tau, T=k\tau,{\cal W}$ are used for
all the lassos (see \cref{ex:vdp0}).
For each source point, the generation of the corresponding lasso stops when the inclusion relation (*) is
verified, which takes around 1000 seconds of CPU time\footnote{which means a total of nearly 30 hours of CPU time for generating the 100 lassos.}.
Note that, as stated by \cref{prop:inclusionbis}, the looping part of each lasso contains an {\em attractive} invariant circle (here, the circle
of centre $(-3,0,0)$ and radius~1, in the $x_2$-$x_3$ plane). %form a ``covering/juxtaposition'' of contiguous lassos. 
%This structure 
%will eventually cover a ``closed surface'' corresponding to the invariant torus whose existence
%is anticipated/(theoretically) guaranteed by the KAM theorem.

\begin{figure}[h!]
\centering
\includegraphics[scale=0.4]{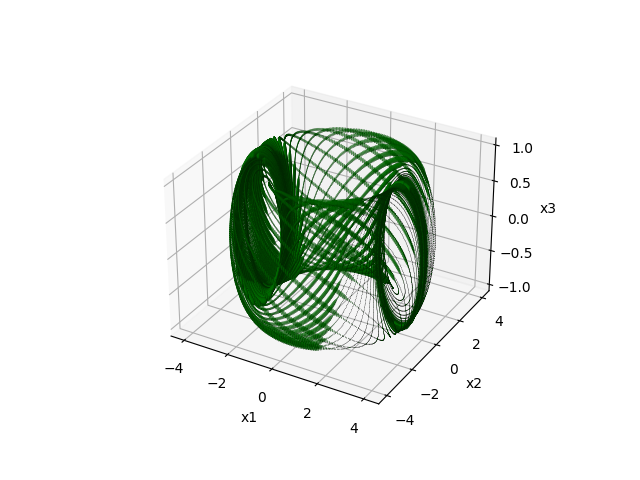}
\includegraphics[scale=0.4]{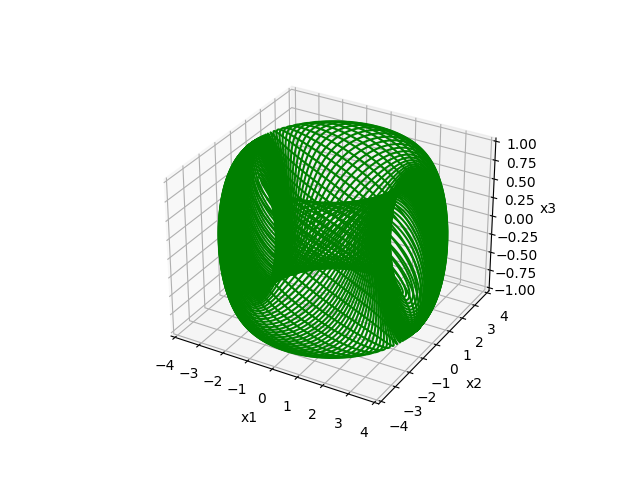}
\caption{{\em Forced VdP}. A set of 50 lassos (top) and 100 lassos (bottom) partially covering
the invariant torus.}
\label{fig:lasso50}
\end{figure}
\end{example}

An other example (coupled VdP oscillators) is given in Appendix.

%\end{example}

\section{Final Remarks}\label{sec:final}

We have introduced a simple technique based on Euler's integration method 
which allows us to construct an invariant structure made of 
a finite number of $n$-dimensional
 balls  covering the invariant torus of the system. 
%In the long version of this paper, we will explain how to formally check that the torus is covered completely.
%The obtained structure is guaranteed to contain the torus and its deformations under the action of small variations of the continuity parameter $\mu$. 
Although it has not been done here, the implementation can be fully 
parallelized since the construction of each lasso is independent of each other.
We have shown on a 3D and a 4D example (one of them close to a torus 
bifurcation) 
how our method gives {\em guaranteed} information on the torus topology. Such a method,
which takes into account the discretization errors,
can help to complement the results obtained with standard numerical methods.

%The same can be done for two coupled VdP oscillators (see \cite{Reichelt2000}).

%\cite{vandenBerg-Queirolo}\cite{Edoh2000} \cite{Hairer99} \cite{Lan2006}\cite{Lanford73} \cite{Rasmussen-Dieci08}\cite{Wang-Slotine05} \cite{Schilder2005}\cite{Krauskopf2000}

%\newpage
%\bibliographystyle{plain}
\bibliographystyle{splncs04}
\bibliography{rp21}
\newpage
\section*{Appendix: Coupled VdP Oscillators Example}
\begin{example}\label{ex:coupled_vdp}
Consider the system of coupled VdP oscillators described in
\cite{Edoh2000}:

\begin{equation}
	\begin{cases}
    \dot{\theta_1} = \beta_1 + \mu\left\lbrace \cos 2\theta_1 - \dfrac{r_2}{r_1}\left[ \sin (\theta_1-\theta_2) + \cos (\theta_1+\theta_2)\right] \right\rbrace +w\\
    \dot{\theta_2} = \beta_2 + \mu\left\lbrace \cos 2\theta_2 - \dfrac{r_1}{r_2}\left[ \sin (\theta_2-\theta_1) + \cos (\theta_1+\theta_2)\right] \right\rbrace +w\\
    \dot{r_1} = r_1(\alpha_1-r_1^2) + \mu\left\lbrace r_1 (1- \sin 2 \theta_1) + Ar_2 \right\rbrace +w\\
    \dot{r_2} = r_2(\alpha_1-r_2^2) + \mu\left\lbrace r_2 (1- \sin 2 \theta_2) + Ar_1 \right\rbrace +w\\
	\end{cases}
\end{equation}
where $A=\sin(\theta_1+\theta_2) - \cos(\theta_1-\theta_2)$.\\
The parameter $\mu$ is the coupling constant and the oscillators decouple for $\mu=0$. Each oscillator has then a unique attractive circle, and the uncoupled product system has a unique attractive invariant torus. The torus persists for a weak coupling and contains two periodic circles, one is attractive and the other is repulsive when $\beta_1=\beta_2$
(see \cite{Edoh2000} for details).
If the manifold $M=(\theta_1,\theta_2,r_1(\theta_1,\theta_2),r_2(\theta_1,\theta_2))$ denotes an invariant torus for the system, then the uncoupled system has an invariant torus defined by $M_1 :=(\theta_1,\theta_2,1,1)$. 
%The attractive circle is defined by the sub-manifold $M_2=(\theta,\theta,1,1)$ where $\theta=\theta_1=\theta_2$. 

Here, we take $w\in{\cal W}=[-0.0001,0.0001]$ and,
as in \cite{Edoh2000}, $\alpha_1=\alpha_2=1.0$, $\beta_1=\beta_2=\beta=0.55$, and $\mu=0.2601$. 
%\footnote{In the case $\beta_1=\beta_2$, the following symmetries can be verified: $r_1(\theta_1,\theta_2)=r_2(\theta_2,\theta_1)$; $r_1(\theta_1,\theta_2)=r_1(\theta_1+\pi,\theta_2+\pi)$; $r_2(\theta_1,\theta_2)=r_2(\theta_1+\pi,\theta_2+\pi)$.???}
%
%Let $\mu=\mu_0+w$\footnote{LF: Est-ce bien sur $\mu$ que $w$ intervient???}\jj{Pour cette exemple, $w$ est  une perturbation additive qui s'applique sur les états (càd: $\dot{\theta_1}+w$, $\dot{\theta_2}+w$ ...)} with 
%
%
Let the time-step $\tau = 10^{-3}$, and the radius of the initial
ball around the source points $\varepsilon=0.1$. 
Let $T=11.425$ be used as an approximation of the exact period $T^*=\frac{2\pi}{\beta}$.%\jj{Je ne suis pas sûr si c'est la bonne valeur de la période exacte du système, j'ai estimé cette valeur à partir des valeurs de $T$,  $\beta_1$ et $\beta_2$} of the periodic solutions.
Each lasso generation now takes around 35 minutes of CPU time.
%Conditions initiales $X(0) = (\theta_1(0), \theta_2(0), r_1(0), r_2(0)) = (0, 0, 1, 1)$ (a completer par differentes valeurs initiales autour du cercle repulsif).
%Cette expérience est avec une perturbation $w=10^{-4}$, un écart initial  $\varepsilon = 0.1$, time-step $\tau = 0.001$. 
We focus visually  on the representation of the projections $r_1(\theta_1,\theta_2)$ and $r_2(\theta_1,\theta_2)$.
%\footnote{The projection  $r_2(\theta_1,\theta_2)$ is similar.}
Ten simulations are thus depicted on \cref{fig:10coupled_vdp_r1,fig:10coupled_vdp_r2},
and the corresponding lassos on \cref{fig:10coupled_bis_vdp_r1,fig:10coupled_bis_vdp_r2}.
The value of $\mu$ is close to the value
$\mu_1\approx 0.2605$ 
for which a torus bifurcation appears
(see \cite{Edoh2000}; cf \cite{Dieci-Bader94,Moore1996}).
This explains the extent of the deformation 
of the  structure, the attractive circle being shaped
like a eight figure on \cref{fig:10coupled_vdp_r1,fig:10coupled_vdp_r2}. %, which lies  ``horizontally'' in the upper  part of 
The source points of the 10 lassos (which coincide with the initial points
of the simulations) have been chosen
close to the repulsive circle (itself estimated by numerical simulation),
as follows:

$X(0)= (0, 3.14159265, 1.05980274, 1.02028354)$

$X(0)= (0.62831853, 3.76991118, 0.95715177, 1.08632695)$ 

$X(0)= (1.25663706, 4.39822972, 1.03960697, 0.93217529)$

$X(0)= (1.88495559, 5.02654825, 0.99657, 1.09545089)$

$X(0)= (2.51327412, 5.65486678, 1.02811851, 1.0178553)$

$X(0)= (3.14159265, 0, 1.08476381, 0.97121437)$

$X(0)= (3.76991118, 0.62831853, 0.97369993, 0.93966289)$ 

$X(0)= (4.39822972, 1.25663706, 1.05513594, 1.00555761)$

$X(0)= (5.02654825, 1.88495559, 0.98407245, 1.09722914)$

$X(0)= (5.65486678, 2.51327412, 0.98484401, 0.93636707)$.\\
For each source point, the inclusion relation (*) is checked for
$i=3,4$ or $5$.

\begin{figure}[h!]
\centering

\includegraphics[scale=0.2]{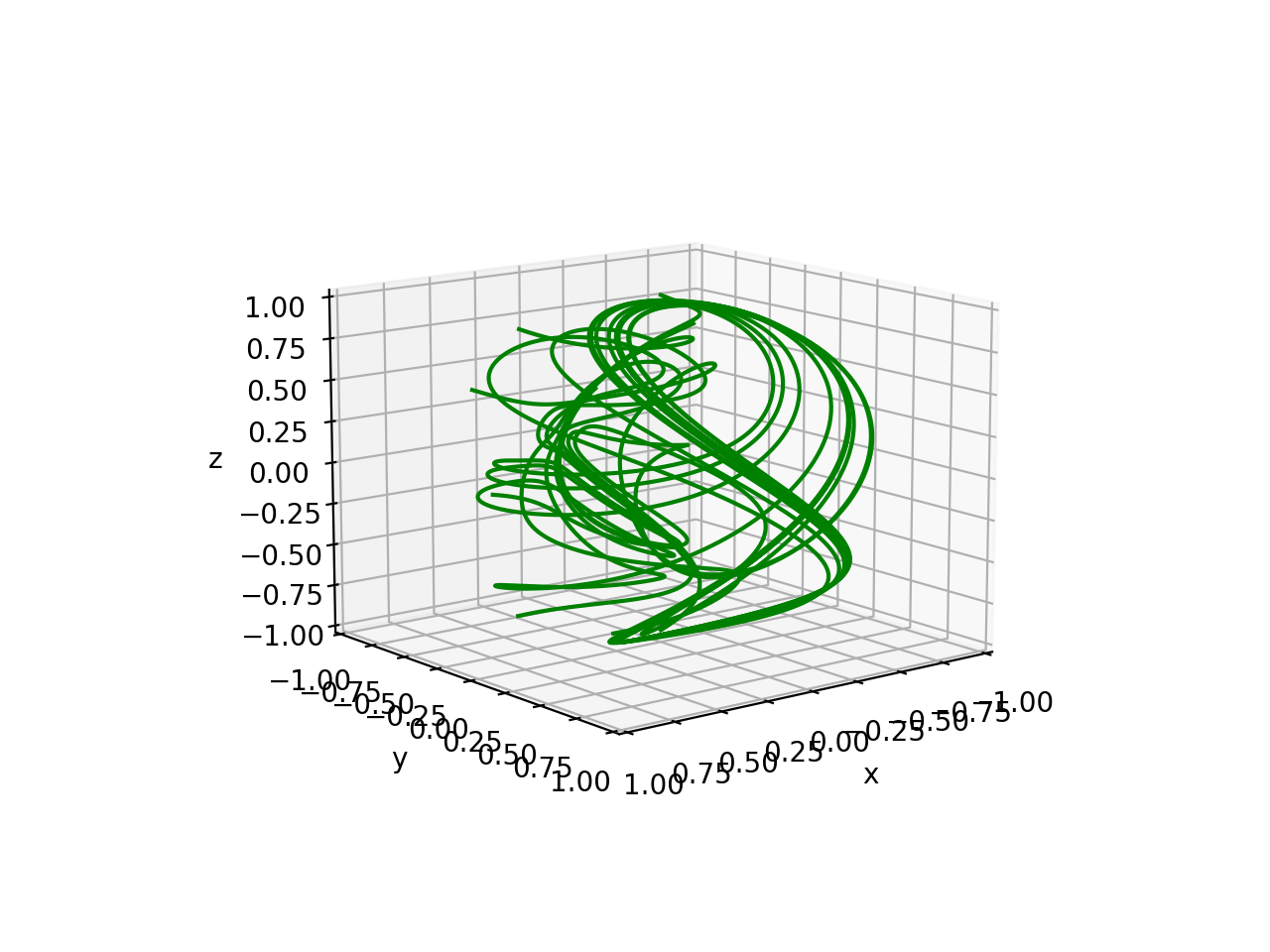}

\includegraphics[scale=0.2]{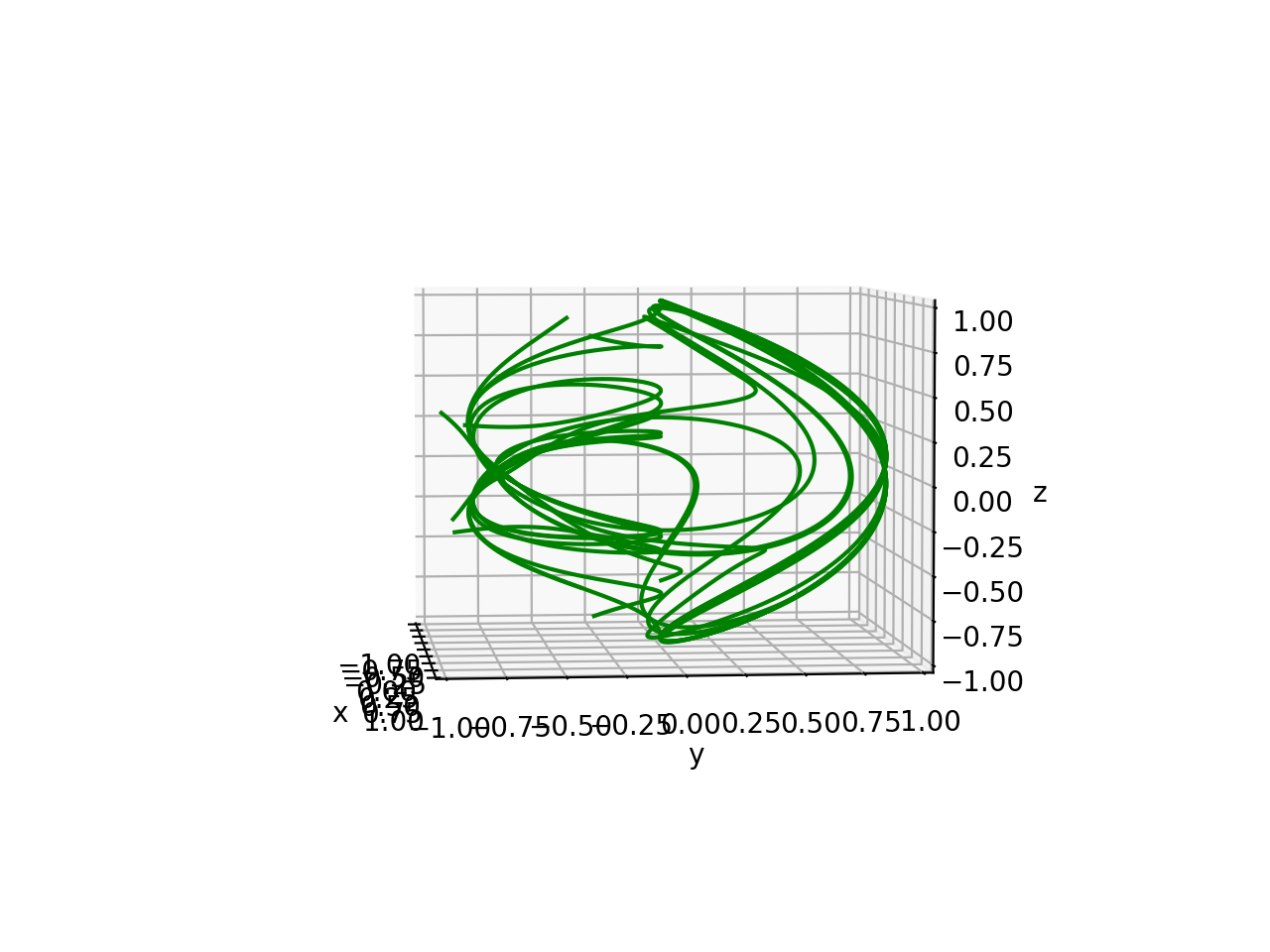}

\caption{{\em Coupled VdP}. The function $r_1(\theta_1,\theta_2)$
corresponding to 10 simulations, under two different views.}
\label{fig:10coupled_vdp_r1}
\end{figure}

\begin{figure}[h!]
\centering
\includegraphics[scale=0.6]{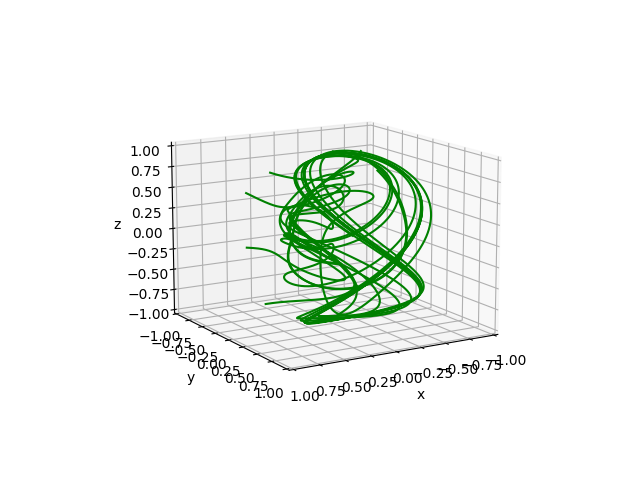}
\includegraphics[scale=0.6]{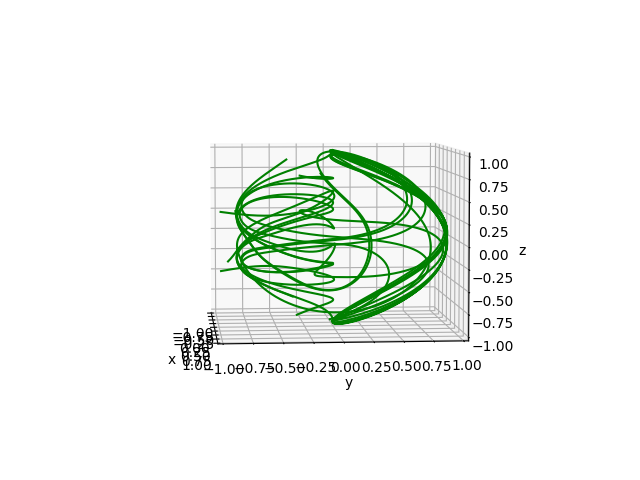}
\caption{{\em Coupled VdP}. The function $r_2(\theta_1,\theta_2)$
corresponding to 10 simulations, under two different views.}
\label{fig:10coupled_vdp_r2}
\end{figure}

\begin{figure}[h!]
\centering
\includegraphics[scale=0.4]{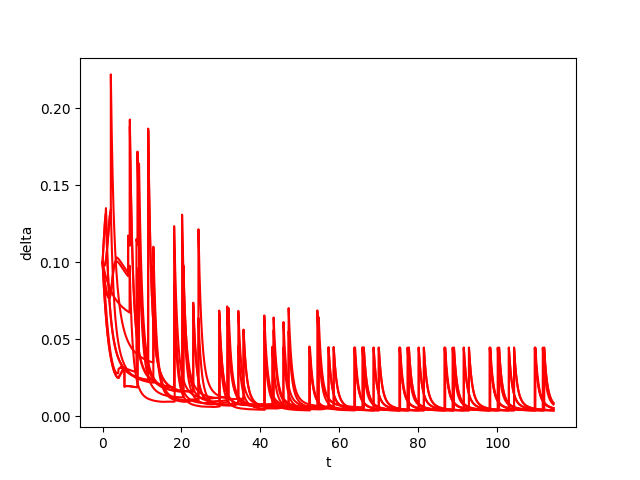}

\includegraphics[scale=0.6]{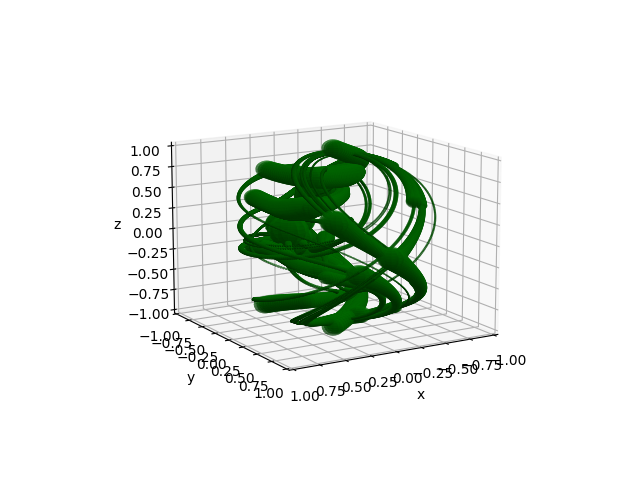}

\includegraphics[scale=0.6]{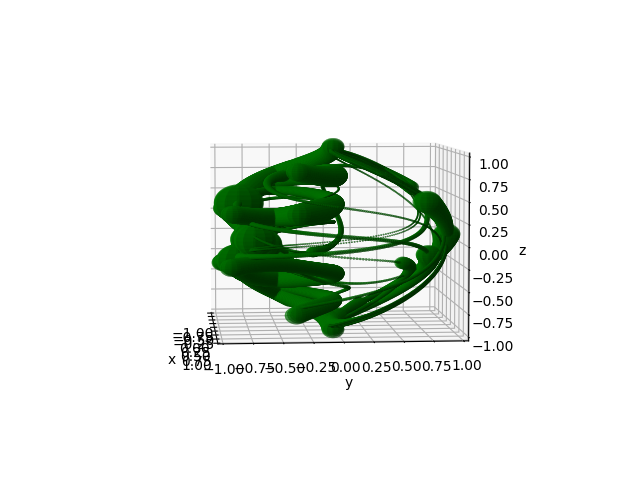}

\caption{{\em Coupled VdP}. Middle and bottom:
The function $r_1(\theta_1,\theta_2)$ corresponding to the lassos associated with the 10 simulations
of \cref{fig:10coupled_vdp_r1},
under the same views. Top: the radius $\delta_{{\cal W}}(t)$ of 
these lassos.}
\label{fig:10coupled_bis_vdp_r1}
\end{figure}

\begin{figure}[h!]
\centering
\includegraphics[scale=0.6]{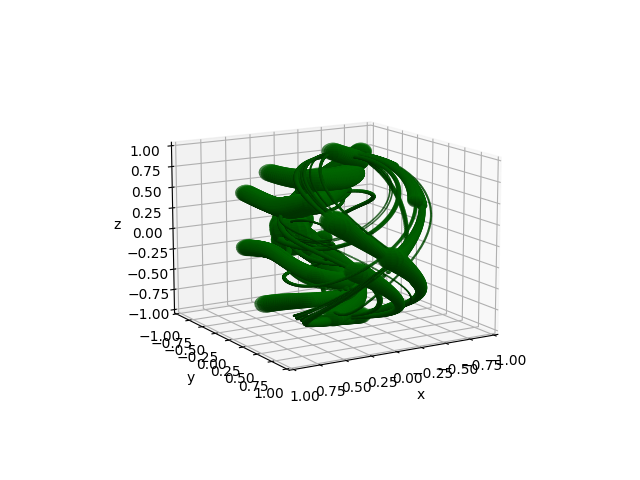}
\includegraphics[scale=0.6]{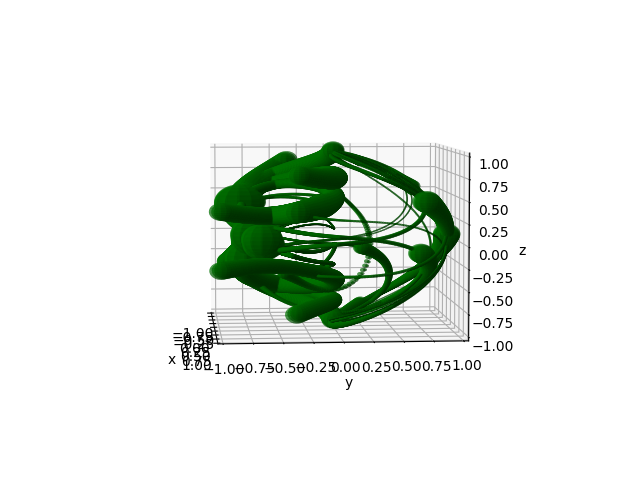}
\caption{{\em Coupled VdP}. Top and bottom:
The function $r_2(\theta_1,\theta_2)$ corresponding to the lassos associated with the 10 simulations
of \cref{fig:10coupled_vdp_r2},
under the same views. }
\label{fig:10coupled_bis_vdp_r2}
\end{figure}

%\begin{figure}[h!]
%\centering
%\includegraphics[scale=0.2]{figures/IMG_2138.PNG}
%\caption{{\em Coupled VdP}. Side view of the torus: attractive circle with a figure-eight shape (with 12 loops???), in the right part;  repulsive circle in the left part (horizontal plane $z=0$).}
%\end{figure}
%\begin{figure}[h!]
%\centering
%\includegraphics[scale=0.4]{figures/IMG_2143.PNG}
%\caption{{\em Coupled VdP}. Top view of the torus: attractive circle in the upper part; repulsive circle in the lower part (vertical plane $x=0$)}
%\label{fig:tori1}
%\end{figure}

%\begin{figure}[h!]
%\includegraphics[scale=0.2]{figures/IMG_2144.PNG}
%\includegraphics[scale=0.2]{figures/IMG_2150.PNG}
%\caption{{\em Coupled VdP}. Tori
%(in both frames, the part corresponding to the torus lies ``horizontally'' at the upper part of the green structure while the 2 ``vertical'' rings at the lower part are 2 attractive circles in formation). }
%\label{fig:tori2}
%\end{figure}

\end{example}

\end{document}